\documentclass[nopacs,twocolumn,preprintnumbers,notitlepage,amsmath,amssymb,superscriptaddress]{revtex4-2}

\usepackage[utf8]{inputenc}
\usepackage[utf8]{inputenc}
\usepackage{physics}
\usepackage{tikz}
\usetikzlibrary{shapes,arrows.meta,decorations.pathreplacing}
\usepackage{mathtools}
\usepackage{blindtext}
\usepackage{ dsfont }
\usepackage{subfiles}
\usepackage{subcaption}
\usepackage[labelfont=bf,
   justification=justified,
   format=plain]{caption} 
\graphicspath{ {./Figures/} }
\usepackage{graphicx}
\usepackage{dcolumn}
\usepackage{bm}
\usepackage{amsmath}
\usepackage{amssymb}
\usepackage{stackengine}
\newcommand{\Lim}[1]{\lim\limits_{#1}}
\newcommand{\Sim}[1]{\mathrel{\mathop{\kern 0pt\sim}\limits_{#1}}}
\newcommand{\To}[1]{\mathrel{\mathop{\kern 0pt\to}\limits_{#1}}}
\newcommand{\Propto}[1]{\mathrel{\mathop{\kern 0pt\propto}\limits_{#1}}}

\newcommand{\dw}{d_\text{w}}

\makeatletter
\AtBeginDocument{\let\LS@rot\@undefined}
\makeatother
\usepackage[]{pdfpages}

\setstackgap{S}{0.5pt}

\begin{document}

\title{Persistence exponents of  self-interacting random walks}
\author{J. Br\'emont}
\affiliation{Laboratoire de Physique Th\'eorique de la Mati\`ere Condens\'ee, CNRS/Sorbonne Université, 
 4 Place Jussieu, 75005 Paris, France}
\affiliation{Laboratoire Jean Perrin, CNRS/Sorbonne Université, 
 4 Place Jussieu, 75005 Paris, France}
\author{L. R\'egnier}
\affiliation{Laboratoire de Physique Th\'eorique de la Mati\`ere Condens\'ee, CNRS/Sorbonne Université, 
 4 Place Jussieu, 75005 Paris, France}
\author{R. Voituriez}
\affiliation{Laboratoire de Physique Th\'eorique de la Mati\`ere Condens\'ee, CNRS/Sorbonne Université, 
 4 Place Jussieu, 75005 Paris, France}
\affiliation{Laboratoire Jean Perrin, CNRS/Sorbonne Université, 
 4 Place Jussieu, 75005 Paris, France}
 \author{O. B\'enichou}
\affiliation{Laboratoire de Physique Th\'eorique de la Mati\`ere Condens\'ee, CNRS/Sorbonne Université, 
 4 Place Jussieu, 75005 Paris, France}
\date{\today}
\begin{abstract}
    The persistence exponent, which characterises the long-time decay of the survival probability of stochastic processes in the presence of an absorbing target, plays a key role in quantifying the dynamics of fluctuating systems. Determining this exponent for non-Markovian processes is known to be a difficult task, and exact results remain scarce despite sustained efforts. In this Letter, we consider the fundamental class of self-interacting random walks (SIRWs), which display long-range memory effects that result from the interaction of the random walker at time $t$ with the territory already visited at earlier times $t'<t$. We compute exactly the persistence exponent for all physically relevant SIRWs. As a byproduct, we also determine the splitting probability of these processes. Besides their intrinsic theoretical interest, these results provide a quantitative characterization of the exploration process of SIRWs, which are involved in fields as diverse as foraging theory, cell biology, and machine learning.
\end{abstract}

\maketitle
The survival probability \( S(x,t) \) is the probability that a stochastic process \( (X_t) \) remains below a given threshold $x$ until time \( t \). This fundamental quantity is related to the statistics of the entire history of the process $(X_t)$, and, as such, plays a crucial role in determining the dynamics of the process. For symmetric, scale-invariant stochastic processes, the long-time behavior of \( S(x,t) \) follows a power-law decay of the form \( S(x,t) \propto t^{-\theta} \), where \( \theta \) is referred to as the persistence exponent. While an exact computation of $S(x,t)$ is usually out of reach, with the exception of a few specific examples of processes, the determination of the exponent \( \theta \) is a more accessible task and has been the focus of a significant body of research; see \cite{bray} for a comprehensive review on the subject. \par 
In the simplest case of $1d$, translation-invariant symmetric Markov processes, the Sparre-Andersen theorem \cite{andersen} yields the universal persistence exponent  \( \theta = 1/2 \).  In contrast, for processes with memory—i.e. non-Markovian processes—the persistence exponent \( \theta \) has been exactly determined exactly only for a few examples of Gaussian processes, such as the fractional Brownian motion (fBM) \cite{molchan}, the diffusive elephant random walk \cite{alex} and the $1d$ voter model \cite{derrida, krapivsky}. 
Approximate determinations of the persistence exponent $\theta$ have been obtained perturbatively for general, weakly non-Markovian Gaussian processes \cite{levernier, bray}. \par
\begin{figure}
    \begin{subfigure}{\columnwidth}
		\includegraphics[width=\textwidth]{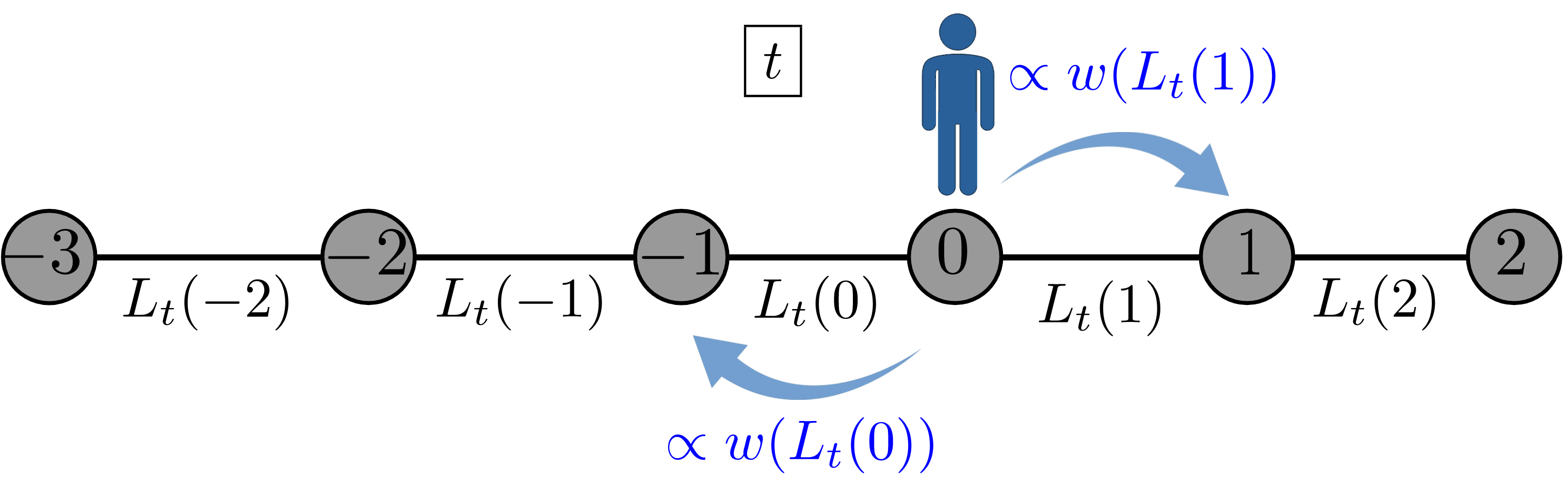}
    \end{subfigure}
    \begin{subfigure}{\columnwidth}
        \includegraphics[width=\textwidth]{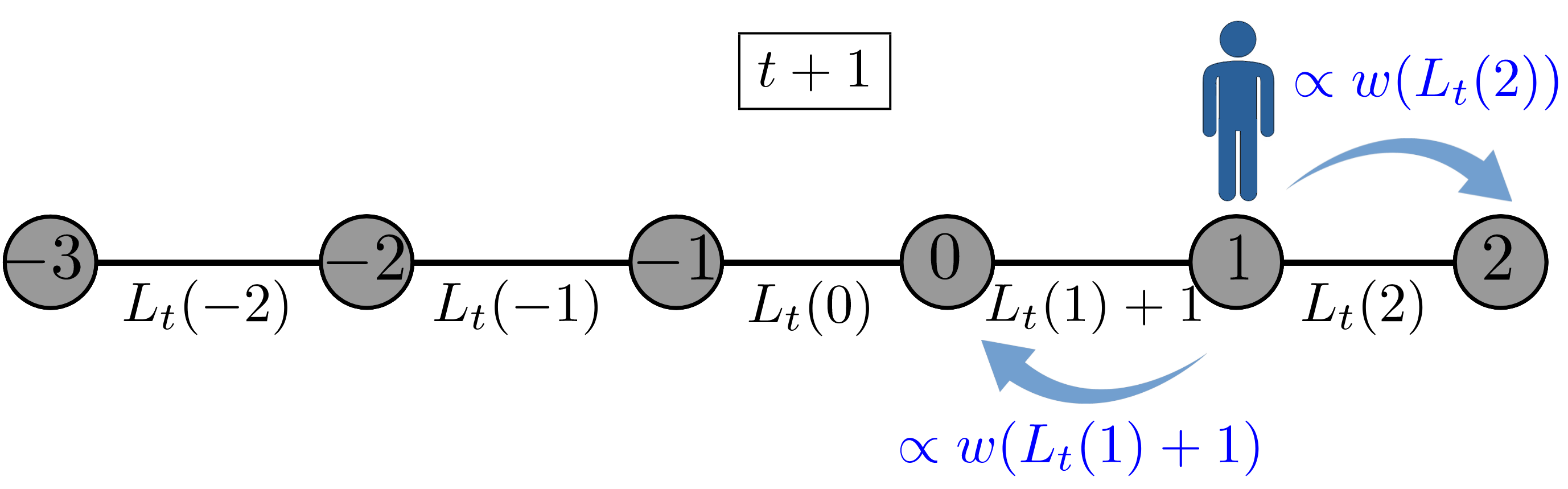}
    \end{subfigure}
    \caption{Sketch of a general SIRW. The sketch on top shows a SIRW standing on site $0$ at time $t$, along with its edge local time profile $(L_t(x))$. Its next jump will be along the edge $\{0,1\}$ : the probability to choose this edge is $\frac{w(L_t(1))}{w(L_t(1)) + w(L_t(0))}$. The lower sketch shows the situation at time $t+1$. The SIRW stands on site $1$, and the edge local time profile has been locally modified on the traversed edge: $L_{t+1}(x) = L_t(x) + \delta_{x,1}$.}
    \label{FIG:dessin}
\end{figure}

 
For processes that are both non-Markovian and non-Gaussian, exact results for $\theta$ are exceedingly rare \cite{nongaussian}. This Letter fills this gap by providing exact values of $\theta$ for the broad class of self-interacting random walks (SIRWs), which are fundamental models of transport. SIRWs are characterized by long-lived memory effects that emerge from the interaction of the random walker at time $t$ with all the territory it has visited at earlier times $t' < t$. More precisely, this model can be defined (see FIG. \ref{FIG:dessin}) as a random walk $X_t$ on $\mathbb{Z}$ with jump probabilities $P_t(x\pm1|x)$ from site $x$ to site $x\pm 1$ given by
\begin{equation}
    P_t(x + 1|x) = \frac{w(L_t(x+1))}{w(L_t(x+1) + w(L_t(x))} = 1-P_t(x-1|x)
\end{equation}
where $L_t(x)$ is the edge local time, defined as the number of times the walker has crossed the unoriented edge $\{x,x- 1\}$ up to time $t$, and $w$ is some positive weight function. Note that  decreasing weight functions $w(n)$ lead to self-repulsion of the walker, while increasing $w(n)$ lead to self-attraction. In fact, as soon as $\sum_{n} w(n)^{-1}$ diverges, the walk is not bounded and $\langle X_t^2 \rangle$ diverges with time as $\langle X_t^2 \rangle \Propto{t \to \infty} t^{2/\dw}$, defining the walk dimension $\dw$ \cite{toth95,toth96}. \par 
The SIRW model has broad applications, as it captures physical situations where a random walker induces persistent, local perturbations in its environment, such as leaving footprints that affect its future dynamics. These self-interactions have been qualitatively observed in various biological contexts, including ants and other animals that leave chemical cues \cite{ants,animals}. More recently, such interactions have been identified in different types of living cells \cite{cellattract,naturealex,lucas}, where the dynamics of these systems were shown to be quantitatively described by the SIRW model. Memory effects in these systems lead to aging and subdiffusion in space exploration \cite{cellattract,naturealex,lucas}. \par
This broad applicability has sparked interest in both mathematics \cite{pemantle,toth96,carmona-yor,diffusive-tsaw,dumaz} and physics \cite{parisi, prasad, sapo, alex, alexprx, abc, structural, grassberger, naturealex, golestanian1,golestanian2,burioni}, with related examples including locally activated random walks \cite{larw-pre} and reinforced walks like the elephant \cite{schutz,baur,bercu}, monkey \cite{boyer} and range-controlled \cite{range-controlled} walks. While, for SIRWs, the walk dimension $\dw$  \cite{davis,toth96,sapo,prasad,diffusive-tsaw} (see Table \ref{tab:parameters} for explicit expressions) and even the propagator \cite{dumaz,satw-prl} have been obtained, the determination of the persistence exponent $\theta$ has remained out of reach, and is at the core of this Letter. \par
Before stating our results, we first remind that the physically relevant cases, corresponding to monotonic weight functions $w(n)$, have been shown to define three different universality classes of SIRWs \cite{toth96}, depending on the large-$n$ asymptotics of $w(n)$. Each class is characterized by the large-scale behaviour of the SIRW, and is independent of the choice of microscopic rules \cite{Note1}. In particular, the exponents $\dw$ and $\theta$ of all models within a given class are the same.
The first class is that of the once-reinforced random walk, also coined in the literature 'self-attracting walk' (SATW$_\phi$), defined by $w_1(n =0) \equiv 1/\phi, w_1(n\geq 1) \equiv 1$. The SIRW belongs to SATW$_\phi$ if $w(n)$ tends to a non-zero limit for $n$ large \cite{Note2}. 
The second class is that of the polynomially self-repelling walk (PSRW$_\gamma$), $w(n) \Sim{n\to \infty} w_2(n) \equiv n^{-\gamma}, \gamma > 0$. The third class is that of the sub-exponential self-repelling walk (SESRW$_{\kappa,\beta}$), $w(n) \Sim{n\to \infty} w_3(n) \equiv e^{-\beta n^\kappa}, \kappa, \beta>0$. Note that this last class contains the seminal SIRW model known as TSAW (true self-avoiding walk) \cite{parisi}, corresponding to an exponential weight function, i.e. the SESRW$_{1,\beta}$ class. \par 
The main result of this Letter is the exact determination of the persistence exponent $\theta$ for each of these universality classes (see FIG. \ref{fig:fpt_densities} for a numerical confirmation)
\begin{equation}
\label{persistence-exp}
    \theta = \begin{cases}
        \frac{\phi}{2} \text{ (SATW$_\phi$)}\\
        \frac{1}{4} \text{ (PSRW$_\gamma$)} \\
        \frac{1}{2} \frac{\kappa+1}{\kappa+2}  \text{ (SESRW$_{\kappa,\beta}$).}
    \end{cases}
\end{equation}
\par 
Let us outline our $2$-step strategy for obtaining \eqref{persistence-exp}. First, relying on Ray-Knight theory \cite{toth96}, we will compute the splitting probability of SIRWs (see \eqref{splitting-result}), which is an important byproduct of our approach. Then, we will use a general scaling relation, obtained in \cite{satya}, relating  persistence exponents to local behaviors of splitting probabilities.\par
\textit{Determination of the splitting probability.}
Define the splitting probability $Q_+(k,-m)$ to be the probability that the SIRW, starting at $0$, hits site $k$ before site $-m$, where $k,m$ are positive integers. The splitting probability is an important quantity by itself \cite{redner}, as it quantifies the likelihood of a particular outcome out of two possibilities for a random process. Beyond the celebrated example of the Gambler's ruin problem, splitting probabilities were found to be useful in the context of population dynamics \cite{popdyn} and helix-coil transition in polymer physics \cite{Oshanin_2009}. Note that very few splitting probabilities have been computed for non-Markovian processes (exceptions can be found in \cite{wiese, dolgushev, rap, satya}). \par
\begin{figure}
    \begin{tikzpicture}[scale=1]
    \draw[->] (-4,0) -- (4,0) node[right] {}; 
    \foreach \x in {-3,-2}{
      \filldraw[gray] (\x,0) circle (2pt);
    }
    \node[below] at (-2,0) {$-2$};
    \foreach \x in {-1,0,1,2,2}{
      \filldraw (\x,0) circle (2pt);
      \node[below] at (\x,0) {$\x$};
    }
    \filldraw (3,0) circle (2pt);
    \draw[thick] (2.85,-0.15) -- (3.15,0.15);
    \draw[thick] (3.15,-0.15) -- (2.85,0.15);
    \node[below] at (-3,0) {$-m$};
    \node[below] at (-2.5,0.3) {$\dots$};
    \node[below] at (0,0) {$0$};
    \node[below] at (3,0) {$k$};
    \draw[decorate,decoration={brace,amplitude=5pt},thick] (-1,0.2) -- (3,0.2) 
    node[midway,above,yshift=0.2cm] {$L+1=5 \text{ sites visited}$};
    \end{tikzpicture}
\caption{Illustration of the observable $q_+(k,L)$ in the case $k=3,L=4$. The sketch represents visited sites (in black), unvisited sites (in gray) as well as the walker's position (crossed), at the time $T_k$ of first arrival to site $k$. In this picture, it is clear that site $-1$, which is the leftmost site visited, was hit before site $k=3$. This exemplifies the decomposition \eqref{splitting} of the splitting probability $Q_+(k,m)$ in terms of the observable $q_+(k,L)$.}
\label{fig:qplus}
\end{figure}
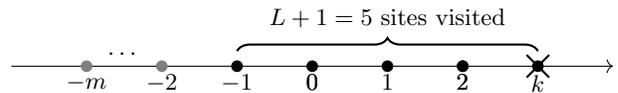

 
We will make use of the probability distribution $q_+(k,L)$ of the number $L+1$ of visited sites by the SIRW at time $T_k$, defined as the first hitting time of site $k$ \cite{satw-prl, klinger}. See FIG. \ref{fig:qplus} for an illustration. If $k$ is reached before $-m$, the number of visited sites ranges from $k+1$ to $k+m$. This observation yields the identity
\begin{equation}
\label{splitting}
    Q_+(k,-m) = \sum_{j=k}^{k+m-1} q_+(k,j).
\end{equation}
The determination of $Q_+(k,m)$ thus reduces to computing $q_+(k,L)$. \par 
Our strategy to determine $q_+(k,L)$ is as follows. The main obstacle to obtaining quantitative results at this point is that the SIRW $X_t$ is highly non-Markovian. For general non-Markovian processes, memory effects are often due to hidden variables that are coupled to the position process $X_t$ and can  have intricate statistics, which renders exact results difficult to obtain. Here, the hidden variables that govern the non-Markovian evolution of the SIRW from time $t$ to time $t+1$ are the set of all edge local times $(L_t(x))_{x\in \mathbb{Z}}$, which correspond to the number of times the edge $\{x,x-1\}$ has been visited up to time $t$. The key points, detailed below, that allow us to obtain explicit results for $q_+(k,L)$ are: (i) this observable can be written in terms of the edge local times $(L_{T_k}(x))_{x\in \mathbb{Z}}$;(ii) in fact, the distribution of $(L_{T_k}(x))_{x\in \mathbb{Z}}$ is known \cite{toth96}. \par 
\begin{figure}
\label{FIG:loctime}
    \begin{subfigure}{0.48\columnwidth}
		\includegraphics[width=\textwidth]{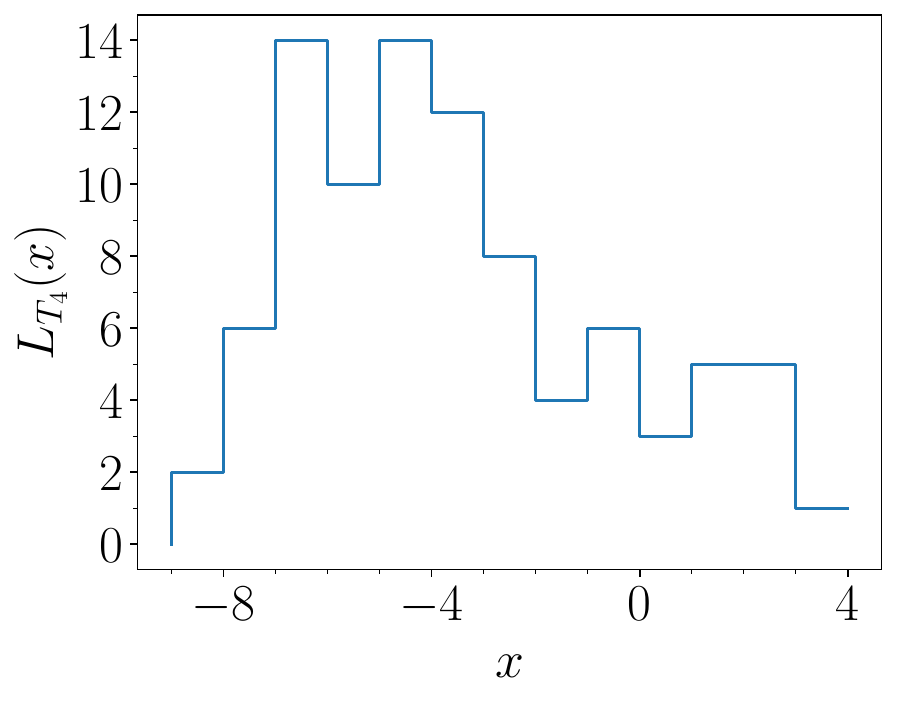}
		\begin{picture}(0,0)
		\put(20,80){(a)}
		\end{picture}
		\label{dec_2d}
		\vspace*{-20pt}
    \end{subfigure}
    \begin{subfigure}{0.48\columnwidth}
        \includegraphics[width=\textwidth]{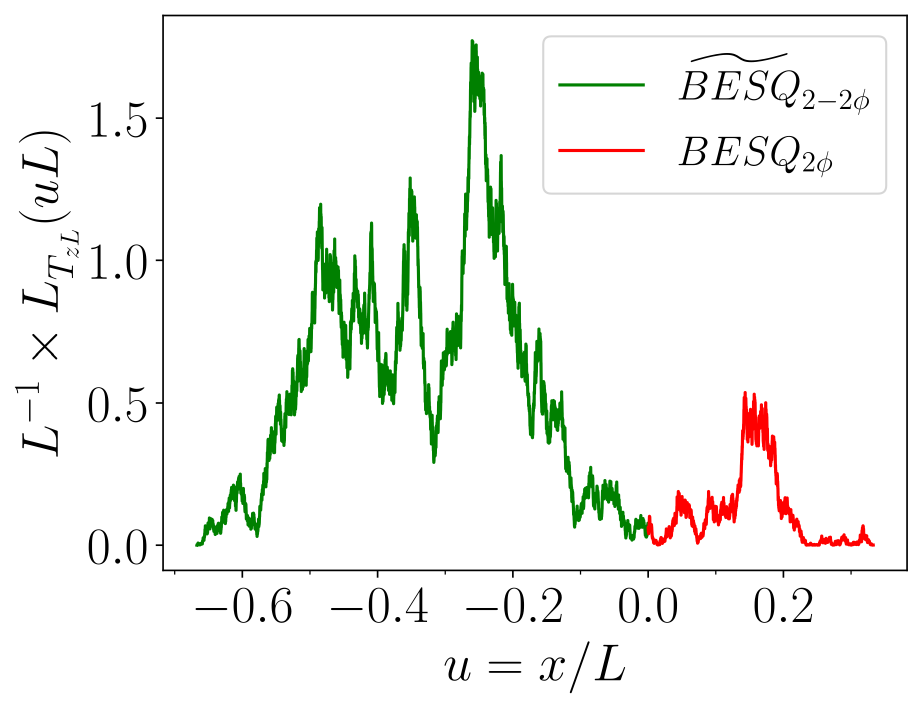}
        \begin{picture}(0,0)
		\put(-30,85){(b)}
		\end{picture}
		\vspace*{-20pt}
    \end{subfigure}
    \caption{(a) Edge local time process $(L_{T_4}(x))_{x\in \mathbb{Z}}$ of a SIRW. The process is stopped at the first hitting time of site $k=4$. Here, we see that the leftmost site visited by the SIRW is $-9$, as $L_{T_4}(x) = 0$ for $x \leq -9$. Thus $\Xi_0 = 9$ as defined in \eqref{q-partition}. A total of $L+1 = 4 + 9 + 1 = 14$ sites have been visited : such a realization contributes to the probability $q_+(k=4,L=13)$. (b) Edge local time process scaled by the number of visited sites $L^z_L(u) = L^{-1}L_{T_{zL}}(Lu)$ of a SATW$_{\phi}$ where $\phi = \log 2$. The limit is defined for $k=zL$ large, and here we chose $k=5000$. The number of sites visited, although random, is inevitably large : $L\sim 15000$ here, and $z = k/L \sim 0.33$ is the fraction of visited sites to the right of $0$. Thus, this realization contributes to the probability $q_+(z\sim 0.33)dz$. Ray-Knight theory \eqref{rayknight-result} states that, as $L\to \infty$,  the red process $(L^z_L(uL))_{0\leq u \leq z}$ converges to $(Y_{2\phi}((z-u)L)_{0\leq u \leq z}$ where $Y_{2\phi}$ is a BESQ$_{2\phi}$ process conditioned on vanishing at $0$. In turn, the green process $(L^z_L (uL))_{0\leq u \leq 1-z}$ converges to $(\tilde{Y}_{2-2\phi}(-u))_{0\leq u \leq 1-z}$ where $\tilde{Y}_{2-2\phi}$ is a $\widetilde{\text{BESQ}}_{2-2\phi}$ process. The red and green processes are independent given their common value at $u=0$.}
    \label{fig:loc_times}
\end{figure}
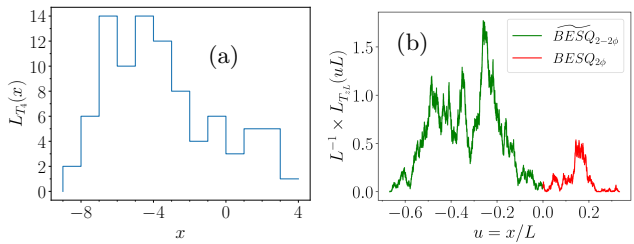

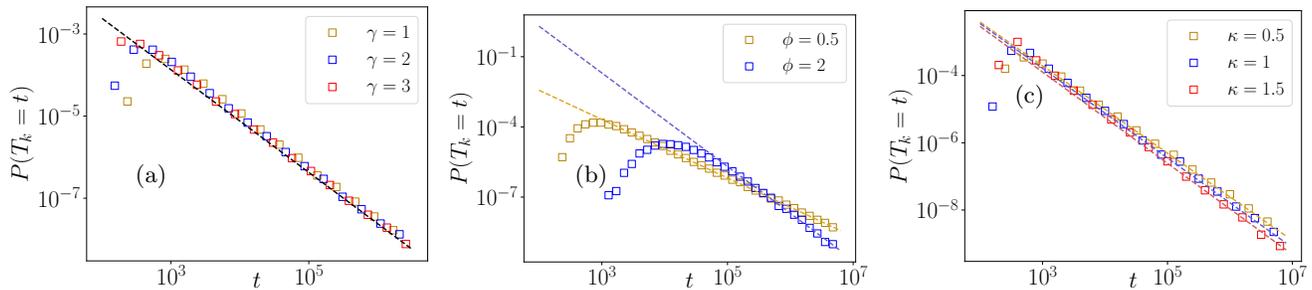
\begin{figure}[t]
    \begin{subfigure}{0.32\columnwidth}
		\includegraphics[width=\textwidth]{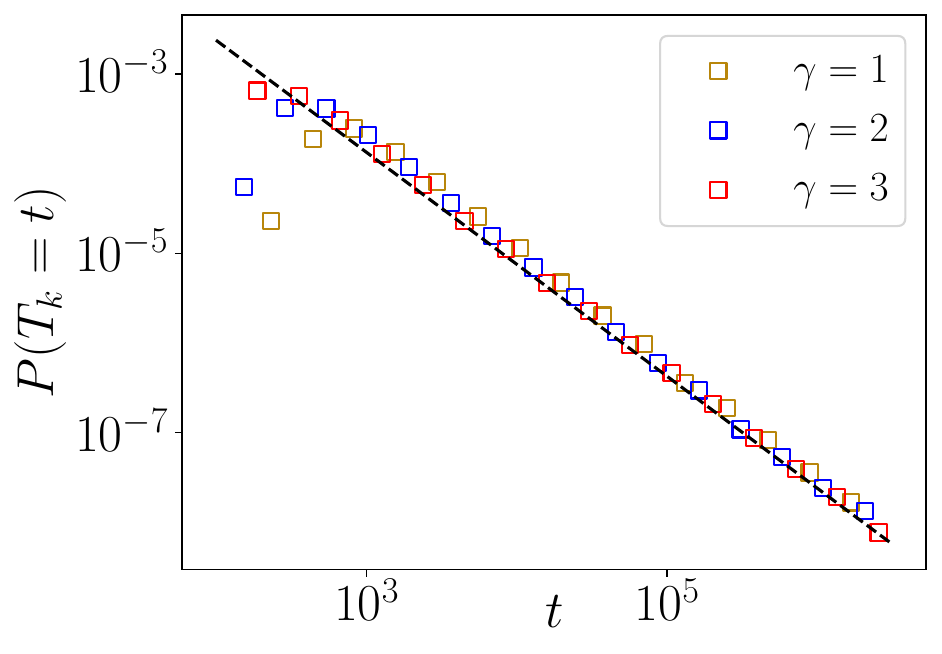}
		\begin{picture}(0,0)
		\put(-30,55){(a)}
		\end{picture}
		\label{dec_2d}
		\vspace*{-20pt}
    \end{subfigure}
    \begin{subfigure}{0.32\columnwidth}
        \includegraphics[width=\textwidth]{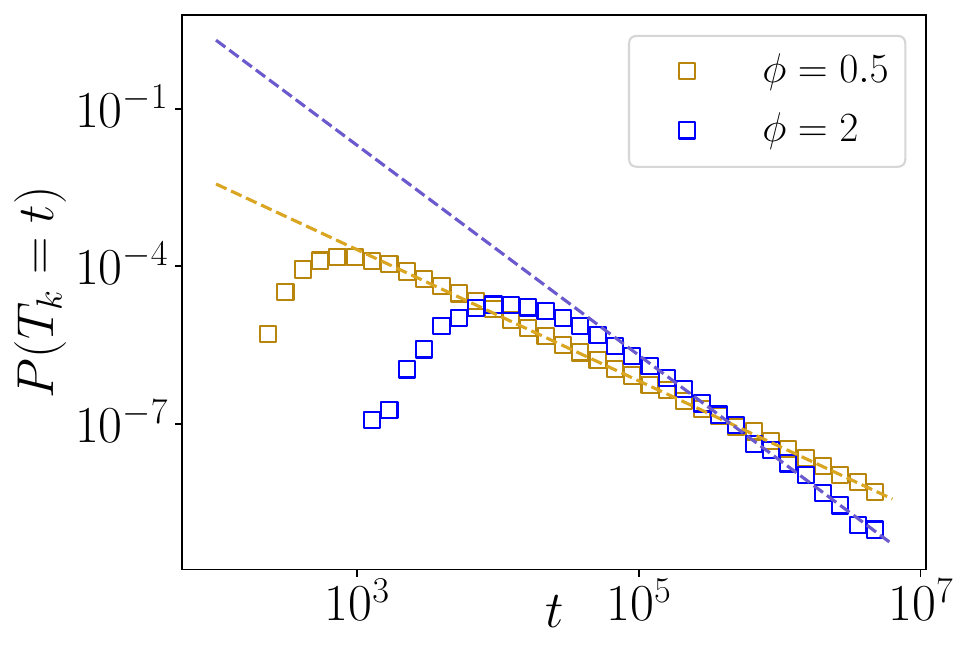}
        \begin{picture}(0,0)
		\put(-30,55){(b)}
		\end{picture}
		\vspace*{-20pt}
    \end{subfigure}
    \begin{subfigure}{0.32\columnwidth}
        \includegraphics[width=\textwidth]{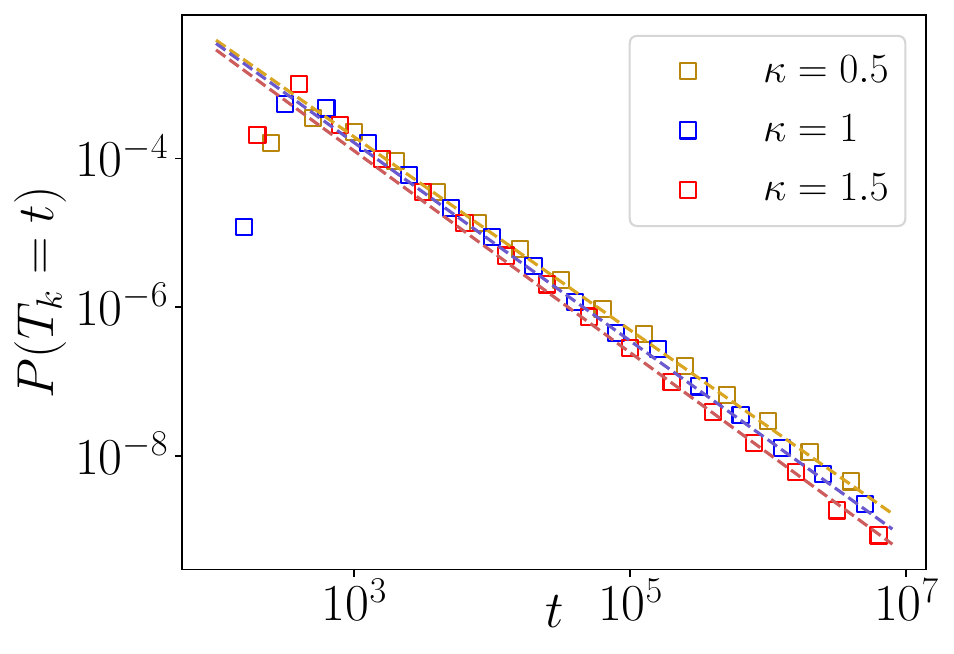}
        \begin{picture}(0,0)
		\put(-30,85){(c)}
		\end{picture}
		\vspace*{-20pt}
    \end{subfigure}
    \caption{Numerical check of \eqref{persistence-exp} performed by analysing the distribution of first-hitting times $T_k$ to site $k=100$, for the three universality classes considered : (a) PSRW$_\gamma$, (b) SATW$_\phi$, (c) SESRW$_{\kappa,\beta=1}$. Symbols represent numerical results while dashed lines represent the theoretical result $P(T_k = t) \Propto{t \to \infty} t^{-1-\theta}$ where $\theta$ is given by \eqref{persistence-exp}.}
    \label{fig:fpt_densities}
\end{figure}

Explicitly, (i) is realized as follows. Let $I_k = \{x \in \mathbb{Z} | L_{T_k}(x) \neq 0\}$ be the support of the edge local time process. Note that we have $L_{T_k}(k)=1$ and $L_{T_k}(x>k)=0$. Because the SIRW is nearest-neighbor and stopped at $T_k$, then necessarily $I_k$ is an interval of the form $I_k = [-m,k]$, where $m$ is a (random and positive) integer. It is easy to see that
\begin{equation}
\label{q-support}
    q_+(k,L) = \mathbb{P}(I_k = [-(L-k)+1,k]) \; .
\end{equation}
Let $-\Xi_0$ be the first negative zero of $(L_{T_k}(x))_{x\in \mathbb{Z}}$ (see FIG. \ref{fig:loc_times} (a) for an illustration). Partitioning over the possible values $a$ of $L_{T_k}(0)$, one finally obtains from \eqref{q-support} the key identity that realizes (i)
\begin{equation}
\label{q-partition}
    q_+(k,L) = \sum_{a=0}^\infty \mathbb{P}(\Xi_0 = L-k|L_{T_k}(0)=a) \mathbb{P}(L_{T_k}(0)=a).
\end{equation}

We now make use of the fact (ii) that the distribution of the edge local time process at inverse local time is known. Let $T_x^h$  be the time at which the SIRW crosses the edge $\{x-1,x\}$ for the $h$-th time \cite{Note3}. The statistics of $(L_{T_x^h}(y))_{y \in \mathbb{Z}}$ are given explicitly by the Ray-Knight theorems for SIRWs, developed by Tòth et al. \cite{toth96} : the process $(L_{T_x^h}(y))_{y \in \mathbb{Z}}$ in the space variable $y$ can be written in terms of squared Bessel processes, defined as follows. When $\delta \in \mathbb{N}$, the squared Bessel process of dimension $\delta$ (BESQ$_\delta$), denoted $Y_\delta(t)$, is simply the norm $\norm{\bold{B}_t}^2$ of a $\delta$-dimensional Brownian motion $\bold{B}_t$. For non-integer values of $\delta$, it is the solution $Y_\delta(t)$ of the stochastic differential equation \cite{yor-bessel} 
\begin{equation}
    dY_\delta(t) = \delta dt + 2\sqrt{|Y_\delta(t)|} dW_t
\end{equation}
where $dW_t$ is uncorrelated Gaussian white noise with variance $1$. More precisely, scaled appropriately, the edge local time process has a continuous limit, explicitly, $ N^{-\alpha}L_{T_{z N}}(Nu) \To{N \to \infty} L^z(u)$, given by \cite{Note4}
\begin{equation}
\label{rayknight-result}
    L^z(u) = \begin{cases} Y_\delta(z-u)^{q}, 0\leq u \leq z \\  \tilde{Y}_{2-\delta}(-u)^{q}, u \leq 0
    \end{cases}
\end{equation}
where $\tilde{Y}_\delta(t)$ is the BESQ$_\delta$ absorbed at $0$, $Y_\delta(0) =0$, and the explicit values of $\alpha, \delta, q$ for each universality class of the SIRW are taken from \cite{tothsurvey} and given in Table \ref{tab:parameters}.
See FIG. \ref{fig:loc_times} (b) for an illustration of this result. 
Importantly, conditioned on their common (random) value $L^z(0)$ at $u=0$, the two processes $(Y_\delta(z-u))_{0\leq u \leq z}, (\tilde{Y}_{2-\delta}(-u))_{u \leq 0}$ introduced in \eqref{rayknight-result} are independent.  In the Supplementary Material (SM) \cite{Note5}, we give a detailed summary of Tóth's articles \cite{toth96,toth95} proving \eqref{rayknight-result} in a physics language.

\begin{table}[h]
\centering 
\begin{tabular}{c|c|c|c}
    & SATW$_\phi$ & PSRW$_\gamma$ & SESRW$_{\kappa,\beta}$ \\
    \hline
    $\alpha$ & 1 & 1 & $(\kappa+1)^{-1}$ \\
    \hline
    $\delta$ & $2\phi$ & 1 & 1 \\
    \hline 
    $q$ & 1 & 1 & $(\kappa+1)^{-1}$\\
    \hline 
    $\dw$ & 2 & 2 & $(\kappa+2)(\kappa+1)^{-1}$ \\
\end{tabular}
\caption{The parameters $\alpha,\delta, q,\dw$ for each class of SIRW.}
\label{tab:parameters}
\end{table}
To fix ideas, in the specific example of TSAW, the scaled edge local time process in the half-space $u<0$, $(L^z(u))_{-\infty<u<0}$, is a simple Brownian motion absorbed at the origin. For $0<u<z$ however, it is a reflected Brownian motion. \par
This gives an explicit answer to (ii) and completes our approach to obtaining an expression for the scaling limit of $q_+(k,L)$, which we define as $q_+(z) = \Lim{L \to \infty} L q_+(k=zL,L)$. In this limit, Eq. \eqref{q-partition} yields, after the change of variables $a \to a^{1/q}$
\begin{equation}
\label{q-fpt-refl}
     q_+(z) = \int_0^\infty  \mathbb{P}(Z_0=1-z|\tilde{Y}_{2-\delta}(0) = a) \mathbb{P}(Y_\delta(0) = a) da
\end{equation}
where $-Z_0=-\Lim{L \to \infty} L^{-1}\Xi_0$ is the first-passage time to $0$ of $\tilde{Y}_{2-\delta}$, and $Y_\delta(0)=0$ by definition. We use the explicit density of $Y_\delta$ \cite{yor-bessel}
\begin{equation}
    \mathbb{P}(Y_\delta(z) = a|Y_\delta(0)=0) = \frac{1}{(2z)^{\delta/2} \Gamma(\delta/2)} a^{\frac{\delta}{2}-1} e^{-\frac{a}{2z}}
\end{equation}
as well as the density of $Z_0$ conditioned on $\tilde{Y}_{2-\delta}(0) = a$
\begin{equation}
   \mathbb{P}(Z_0=1-z|\tilde{Y}_{2-\delta}(0) = a) = \frac{e^{-\frac{a}{2(1-z)}}}{(1-z)\Gamma(\delta/2)} \left(\frac{a}{2(1-z)}\right)^{\delta/2}. 
\end{equation}
This yields 
\begin{equation}
    q_+(z) = \frac{ (z(1-z))^{-\delta/2}}{2^{\delta}(1-z)\Gamma(\delta/2)^2} \int_0^\infty e^{-\frac{a}{2z} - \frac{a}{2(1-z)}} a^{\delta-1} da,
\end{equation}
and finally
\begin{equation}
    q_+(z) = \frac{1}{B(\delta/2)} \frac{(z(1-z))^{\delta/2}}{1-z}
\end{equation}
where $B(a) = \int_0^1 (u(1-u))^{a-1} du$ is a Beta function. From the values of $\delta$ in Table \ref{tab:parameters} for each universality class, this yields 
\begin{align}
\label{q-result}
    q_+(z) = \begin{cases} 
        \frac{1}{B(\phi)} \frac{(z(1-z))^\phi}{1-z}  & (\text{SATW}_\phi) \\ 
        \frac{1}{B(1/2)} \sqrt{\frac{z}{1-z}} & \left( \stackanchor{PSRW$_\gamma$}{SESRW$_{\kappa,\beta}$} \right)
    \end{cases}
\end{align}

As a last step, we obtain the splitting probability from \eqref{q-result}. Its scaling limit writes 
\begin{equation*}
    Q_+(z) = \Lim{L \to \infty} Q_+(k=zL, -m=-(1-z)L)
\end{equation*} 
Using \eqref{splitting}, we have $Q_+'(z) = -q_+(z)/z$, hence $Q_+(z) = 1- \int_0^{z} (q_+(u)/u)du$ where we used that $Q_+(0)=1$. From \eqref{q-result}, we then obtain the splitting probability, which reads
\begin{align}
    Q_+(z) = \begin{cases} I_{1-z}(\phi)  & (\text{SATW}_\phi) \\ I_{1-z}\left(\frac{1}{2}\right) = \frac{2}{\pi} \arcsin{\sqrt{1-z}} & \left( \stackanchor{PSRW$_\gamma$}{SESRW$_{\kappa,\beta}$} \right)
    \end{cases}
    \label{splitting-result}
\end{align}
where $I_z(a) = B(a)^{-1} \int_0^z (u(1-u))^{a-1} du$ is the regularised incomplete Beta function.

This exact result, illustrated in SM, calls for several comments. (i) Let us stress that, even if presented here as a technical step towards the determination of the persistence exponent, the explicit expression of the splitting probability for SIRW constitutes an important result in itself. (ii) Note that the result for SATW$_\phi$ was already computed in \cite{doney} using a completely different method. The arcsine law for the other two classes is new to the best of our knowledge and striking: the splitting probability does not depend on any of the parameters $\gamma,\kappa,\beta$. (iii) The ubiquity of $I_z$ in the expression of the splitting probability of several $1d$ scale-invariant processes was already noted in \cite{satya}, where it was shown to appear for Lévy flights, the random acceleration process, and Sinai diffusion. However, in this article, it was also shown that expressions of splitting probabilities involving $I_z$ do not hold for all non-Markovian processes, as was checked on numerical grounds for the fBM, a paradigmatic non-Markovian process (see also \cite{wiese} for analytical confirmation). In contrast, we show here that $I_z$ is involved in the splitting probability for a whole class of strongly non-Markovian processes, namely, SIRWs. This considerably extends the range of validity of the "universal" expression of the splitting probability in terms of $I_z$. (iv) We give here a new probabilistic interpretation of the integrand $(u(1-u))^{a-1}$ involved in the splitting probability $I_z(a)$, namely, as the term $q_+(u)/u$, hinting at the fundamental nature of this observable. In other words, it provides a physical interpretation of the universal form noted in \cite{satya}.\par
\textit{The persistence exponent from the splitting probability.} In \cite{satya}, it was shown that the splitting probability of a generic symmetric, scale-invariant $1d$ stochastic process $X_t$ starting at $0$  verifies 
\begin{equation}
\label{persistence-from-splitting}
    Q_+(z) \Propto{z \to 1} (1-z)^{\Phi}
\end{equation}
where $\Phi = \dw \theta$, $\dw$ is the walk dimension and $\theta$ is the persistence exponent of $X_t$. 
For the sake of completeness, we give a brief argument that leads to \eqref{persistence-from-splitting}, similar in spirit to the one presented in \cite{levernierUniversalFirstpassageStatistics2018}. 
Let $F(1-z,t)$ be the first-passage time density of $X_t$ to $-(1-z)$ in semi-infinite space. Thanks to the scale-invariance of $X_t$, we can write $F(1-z,t) = f(t/(1-z)^{\dw})/t$, with $f$ a process-dependent scaling function. $X_t$ reaches $-(1-z)$ within a typical timescale $(1-z)^{\dw}$, which, as $z \to 1$, is much smaller than the timescale $z^{\dw}$ needed to reach $z$. Thus, as $z \to 1$, the probability of reaching $z$ before $-(1-z)$ is determined mostly by the contribution of trajectories that reach $-(1-z)$ only after an atypically large time $T>z^{\dw}$, i.e. $Q_+(z) \Propto{z \to 1} \int_{z^{\dw}}^\infty F(1-z,t) dt = \int_{(z/(1-z))^{\dw}}^\infty (f(u)/u) du$. By definition of the persistence exponent $\theta$, we have $f(u) \Propto{u \gg 1} u^{-\theta}$. This yields $Q_+(z) \Propto{z\to 1} (1-z)^{\dw \theta}$, which is exactly \eqref{persistence-from-splitting}. \par 
We now apply \eqref{persistence-from-splitting} to the scale-invariant, symmetric process that is the scaling limit of SIRW considered here. We remind that $\dw$ is given in Table \ref{tab:parameters} and $\Phi$ is given by \eqref{splitting-result}. This finally yields the persistence exponents of SIRW  \eqref{persistence-exp}. \par 
Finally, let us comment on this main result Eq. \eqref{persistence-exp}. (i) This result is exact and confirmed numerically in FIG. \ref{fig:fpt_densities}. (ii) We recover the persistence exponent of the SATW$_\phi$ class that was obtained using a different method in \cite{alex}. (iii) In \cite{levernier}, it was predicted that for processes with stationary increments, one has $\theta = 1-\dw^{-1}$. See \cite{krug,molchan} in the case of the fBM. We indeed find that this holds only for the TSAW and Brownian motion, which are the only two SIRWs with stationary increments \cite{tsrm}. 
\par 
Finally, making use of Ray-Knight theory, we obtained an exact determination of both the splitting probability and the persistence exponent for the broad class of SIRWs. These exact results provide benchmarks for the future analysis of more realistic models of self-interacting processes, which are relevant to model various experimental situations. 
\begin{acknowledgements}
We thank P. Pineau for careful reading of the SM.
\end{acknowledgements}
\onecolumngrid

\twocolumngrid



\section*{Supplementary material}

# Supplementary Material - Persistence exponents of one-dimensional self-interacting random walks

October 24, 2024

For all universality classes of SIRW, Ray-Knight theory gives the joint distribution of all local times when stopping the walk at a given inverse local time. In the following, we give a summary of the content of [1] on Ray-Knight theorems for self-interacting random walks. Seeing how this tool allows us to compute several nontrivial quantities of interest in physics, as illustrated in the main text, we believe that exposing this theory and its inner workings in a language familiar to physicists is interesting in itself. This is the goal of the following sections $1, 2, 3$ of this Supplementary Material.

## 1 General results useful in the following

Define two weight functions $r(n), b(n)$ to be interpreted as the weight given to red (resp. blue) balls of a generalized Polya urn. The number of red (resp blue) balls after $i$ draws is $\rho_i$ (resp $\beta_i$). The transition probabilities of this generalized Polya urn write

$$P((\rho_{i+1}, \beta_{i+1}) = (k+1, l)|(\rho_i, \beta_i) = (k, l)) = \frac{r(k)}{r(k) + b(l)} \quad (1)$$

A crucial quantity is the time $\tau_m$ at which the $m$th red ball is drawn, and the (random) number $\mu(m) = \beta_{\tau_m}$ of blue balls drawn at time $\tau_m$. The following identity, valid for $\lambda < \min_{j \leq m-1} r(j)$, is key to the analysis

$$\mathbb{E}\left[\prod_{j=0}^{\mu(m)-1}\left(1 + \frac{\lambda}{b(j)}\right)\right] = \prod_{j=0}^{m-1}\left(1 - \frac{\lambda}{r(j)}\right)^{-1} \quad (2)$$

We give a short proof by recurrence on $m$ of (2). It can be checked that it holds for $m = 0, 1$. Suppose (2) holds for all weight functions $b, r$ and for all indices $m \leq M$. A simple renewal argument allows us to write $\mu(M+1) = \mu(M) + \tilde{\mu}(1)$, where $\tilde{\mu}(1)$ is the number of blue balls drawn before the first red ball is drawn in a Polya urn with weight functions $\tilde{r}(j) = r(m+j), \tilde{b}(j) = b(j+\mu(M))$. Applying (2) at steps $M$ and 1 then completes the recurrence at step $M+1$.

## 2 The local time process as a network of generalized Polya urns

In this section, we describe the local time process $(L_{T_{k,m}}(x))_{x \in \mathbb{Z}}$ of a self-interacting random walk (SIRW) starting at 0 stopped at the time $T_{k,m}$ defined by

$$T_{k,m} \equiv \inf\{t \geq 0 | \text{the oriented edge } (k-1, k) \text{ has been crossed } m+1 \text{ times up to time } t\} \quad (3)$$

Notice that the SIRW stopped at $T_{k,m}$ necessarily ends at $k$, and has crossed the non-oriented edge $\{k-1, k\}$ a total of $m+1+m = 2m+1$ times. We will focus on the case $k > 0$. We begin by putting a generalized Polya urn at each integer: the index $l$ labels the urn at site $k-l$. The color of the ball drawn at urn $l$ is identified with the direction of the step taken by the RWer at site $k-l$. Our convention, based on the direction of propagation of local times from the source site $k$ (ie, $l = 0$), is the following

$$\begin{cases} l \in ]-\infty, 0] : (\text{jump left}, \text{jump right}) \\ l \in ]0, \infty[ : (\text{jump left}, \text{jump right}) \end{cases} \tag{4}$$

The weight functions $r^l(\rho), b^l(\beta)$ of urn $l$, which represent the weight of drawing a red (resp blue) ball when $\rho$ red balls (resp $\beta$ blue balls) are inside the urn, are defined via the weight function $w(n)$ of the self-interacting random walk. For the weight of drawing a red ball when there are $\rho$ red balls in the urn $l$,

$$r^l(\rho) = \begin{cases} w(2\rho + 1), l \in ]-\infty, 0] \cup [k, \infty[ \\ w(2\rho), l \in [1, k-1] \end{cases} \tag{5}$$

and for the weight of drawing a blue ball,

$$b^l(\beta) = \begin{cases} w(2\beta), l \in ]-\infty, 0] \cup [k, \infty[ \\ w(2\beta + 1), l \in [1, k-1] \end{cases} \tag{6}$$

We give a motivation for (5) and (6). Let us look at the urn $l$ on site $j = k - l < 0, l \in ]k, \infty[$ (see Fig. 1 for an illustration). For the urn $l$, according to (4), red (resp blue) balls represent crossings of $(k-l, k-l+1)$ (resp of $(k-l, k-l-1)$). The RWer on site $k-l$ has necessarily crossed the non-oriented edge $\{k-l, k-l+1\}$ an *odd* number of times, since it will end at site $k$ at time $T_{k,m}$ and $k - l < 0$. Therefore, the probability of choosing the right edge $(k-l, k-l+1)$ for the $x$th time is proportional to $w(2x-1)$. On the contrary, the probability of choosing the left edge $(k-l, k-l-1)$ for the $y$th time is proportional to $w(2y)$.

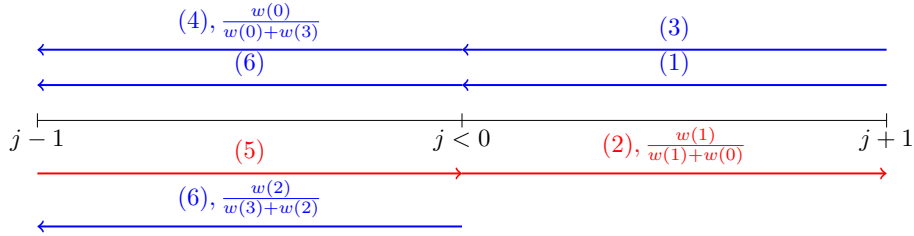

Figure 1: Definition of the Polya urn at a site $j < 0$. Each arrow is accompanied with a labeling $(x)$ that represents time, and if the arrow departs from site $j$, the probability of jumping along this edge while on site $j$ is shown. The color of the arrow is the color of the ball drawn from the urn at the departing site and is defined by (4).

As an example, in Fig. 1, we have that when the RWer steps on the right at (2) where it has not made any departure from $j$ ($\rho = 0$, $\beta = 0$) up to this point, there is one crossing of $\{j, j+1\}$ but no crossing of $\{j-1, j\}$, such that the weight to go to the right is $w(1)$ and to the left $w(0)$. When the RWer steps on the left at (6) where it has made one departure to the right at (2) and one to the left at (4) ($\rho = 1$, $\beta = 1$), there has been 3 crossings of $\{j, j+1\}$ and 2 crossings of $\{j-1, j\}$, such that the weight to go to the right is $w(3)$ and to the left $w(2)$.

Let $m \geq 1$. We define $S_{k,m}^>(l)$ to be the number of times the RWer starting from site 0 at time 0 crossed the oriented edge $(k-l, k-l-1)$, at the time $T_{k,m}$ of $m+1$th arrival to $k$ coming from the left. As we will see, this is a Markov process in the spatial variable $l$, and can be represented as a network of Polya urns. We have, by definition, $S_{k,m}^>(0) = m$. Of course, $S_{k,m}^>(l)$ behaves differently for $l$ in any of the three intervals $]-\infty, 0[, ]k, \infty[$ and $[0, k]$, since the RWer visited every site in $[0, k]$ at least once, but no such condition is imposed on the other intervals. This observation leads to the following representation for $l \geq 1$ (see definition of the random variable $\mu$ in Sec. 1):

$$S_{k,m}^>(0) = m, \quad S_{k,m}^>(l) \equiv \begin{cases} \mu^l \left( S_{k,m}^>(l-1) + 1 \right), l = 1, 2, \ldots, k \\ \mu^l \left( S_{k,m}^>(l-1) \right), l \geq k+1, \ldots \end{cases} \quad (7)$$

For $l \leq 0$, we define

$$S_{k,m}^>(l-1) \equiv \mu^l \left( S_{k,m}^>(l) \right) \quad (8)$$

These definitions are illustrated FIG. 2.

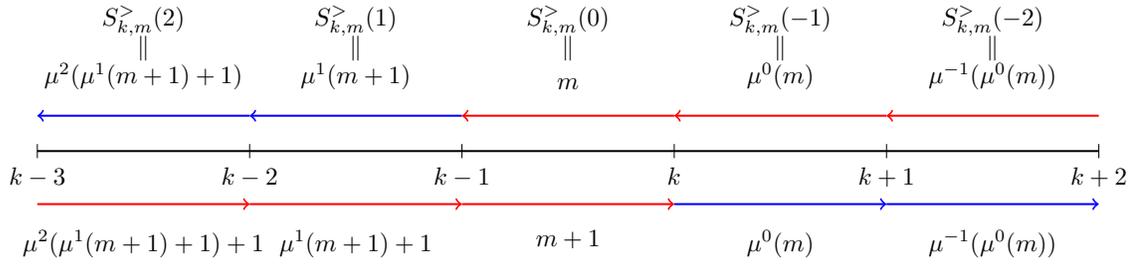

Figure 2: Sketch of the local time process $S_{k,m}^>(l)$ at time $T_{k,m}$, with $k \geq 3$ here. The numbers next to the left (resp right) arrows represent the number of leftwards (resp rightwards) crossings of the associated edge. The colors of the arrows represent the color of the balls associated to their direction in the urn they point out of. Notice the equality of rightwards and leftwards crossings for the edges in the halfspace $x > k$. This is due to 2 facts : i) we start from site 0, ii) we stop the process when $k$ has been visited $m+1$ times from the left. Edges $(x, x+1), x \in [0, k-1]$ have been visited once more than their counterparts $(x+1, x)$ for the same reason. There is always one red and one blue arrow pointing out of each site.

This discussion invites us to define two Markov processes. The first one, $\mathcal{Z}(l)$, defined by $\mathcal{Z}(l+1) = \mu^{l+1}(\mathcal{Z}(l) + 1)$, we will identify with $S_{k,m}^>(l)$ in the interval $l \in [1, k-1]$. According to Eqs. (5)-(6) and FIG. 2, to make this identification, the Polya urns associated to $\mu^l$ must have weight functions $r(j) = w(2j), b(j) = w(2j+1)$. The second Markov process, $\tilde{\mathcal{Z}}(l)$, defined by $\tilde{\mathcal{Z}}(l+1) = \tilde{\mu}^{l+1}(\tilde{\mathcal{Z}}(l))$, we will identify with $S_{k,m}^>(l)$ for $l \in \mathbb{Z}_-$. The Polya urns associated to $\tilde{\mu}^l$ have weight functions $\tilde{r}(j) = w(2j+1), \tilde{b}(j) = w(2j)$.

It now remains to compute the distribution of these processes in the continuous limit. In the following, we compute only the distribution of $\mathcal{Z}(l)$, as the computation for $\tilde{\mathcal{Z}}(l)$ is similar.

3

# 3 Asymptotic distributions of the process $\mathcal{Z}$

Thinking of applying (2), we define

$$U_p(n) = \sum_{j=0}^{n-1} w(2j)^{-p}, V_p(n) = \sum_{j=0}^{n-1} w(2j+1)^{-p} \tag{9}$$

as well as the new Markov chain $\mathcal{Y}(l) = V_1(\mathcal{Z}(l))$. Define $F(x) \equiv \mathbb{E}\left[\mathcal{Y}(l+1)|\mathcal{Y}(l) = x\right] - x = U_1(V_1^{-1}(x) + 1) - x$, where the last equality stems from (2). Finally, again thanks to (2), defining $G(x) \equiv U_2(V_1^{-1}(x) + 1) + \mathbb{E}\left[V_2(V_1^{-1}(Y(l+1)))|Y(l) = x\right]$ allows us to write the increments $\epsilon_l$

$$\mathcal{Y}(l+1) - \mathbb{E}\left[\mathcal{Y}(l+1)|\mathcal{Y}(l)\right] \equiv \sqrt{G(\mathcal{Y}(l))}\epsilon_l \tag{10}$$

such that the $\epsilon_l$ have mean zero, unit variance and (since $\mathcal{Y}$ is Markov) are independent from each other: in the scaling limit, we will interpret $\epsilon_l$ as uncorrelated Gaussian white noise. We thus have

$$\mathcal{Y}(l+1) - \mathcal{Y}(l) = \sqrt{G(\mathcal{Y}(l))}\epsilon_l + F(\mathcal{Y}(l)) \tag{11}$$

It now remains to compute the asymptotics of the functions $F, G$. As is clear from the definitions of $F, G, U_p, V_p$, assuming that $\sum_n w(n)^{-1}$ diverges, the scaling limit of (11) depends only on the asymptotics of $w(n)$, justifying the existence of the universality classes described in the introduction of the main text.

## 3.1 SATW$_\phi$

In this simplest case, we have $U_p(n) = \phi^p + (n-1)$, while $V_p(n) = n$. This yields $F(x) \underset{x \gg 1}{\sim} \phi$. For $G$, given that $U_p(n), V_p(n) \sim n$, we write

$$\begin{aligned} G(x) &= U_2(V_1^{-1}(x)+1) + \mathbb{E}\left[V_2(V_1^{-1}(\mathcal{Y}(l+1)))|\mathcal{Y}(l) = x\right] \sim x + \mathbb{E}\left[\mathcal{Y}(l+1)|\mathcal{Y}(l) = x\right] \\ &= x + U_1(V_1^{-1}(x)+1) \sim 2x \end{aligned} \tag{12}$$

The scaling limit is defined by $Y(t) = \lim_{A \to \infty} A^{-1}\mathcal{Y}(\lfloor At \rfloor)$. Writing $dW_t = \lim_{A \to \infty} A^{-1/2}\epsilon_{\lfloor At \rfloor}$ where $dW_t$ is uncorrelated Gaussian white noise, (11) and the asymptotics of $F, G$ then yield the stochastic differential equation (SDE)

$$dY(t) = \phi dt + \sqrt{2Y(t)}dW_t \tag{13}$$

Hence $Y(t)$ is a BESQ$_{2\phi}$, and, since $\mathcal{Z}(l) = V_1^{-1}(\mathcal{Y}(l)) = \mathcal{Y}(l)$, the limit process $Z(t) = \lim_{A \to \infty} A^{-1}\mathcal{Z}(\lfloor At \rfloor)$ is a BESQ$_{2\phi}$ as well.

## 3.2 PSRW$_\gamma$

### 3.2.1 Asymptotics of $U_1, V_1$ from the Euler-MacLaurin formula

In Tóth's original work [1], a misprint exists in the asymptotic forms of $U_1, V_1$ given for the PSRW. We correct this misprint here. Recall the Euler-MacLaurin formula (at order 1, we will not need more accuracy)

$$\sum_{k=0}^{n} f(k) = \int_0^n f(x)dx + \frac{f(n) + f(0)}{2} + \int_0^n f'(x)(x - \lfloor x \rfloor - 1/2)dx \tag{14}$$

4

Note that the weight function can be defined up to a multiplicative constant. For convenience, we use the weight function $w(n)^{-1} = 2^{-\gamma}(\gamma + 1)n^\gamma$. From the definitions of $U_1, V_1$, (14) yields

$$U_1(n) = n^{\gamma+1} - \frac{\gamma+1}{2}n^\gamma + o(n^\gamma), V_1(n) = n^{\gamma+1} + o(n^\gamma) \tag{15}$$

### 3.2.2 Asymptotics of $F, G$

From the asymptotics of $V_1$, we have $V_1^{-1}(x) = x^{\frac{1}{\gamma+1}} + O(1)$. Thus, writing $n = V_1^{-1}(x)$ and using (15), one has

$$F(x) = U_1(V_1^{-1}(x) + 1) - x = U_1(n+1) - V_1(n) \sim \frac{\gamma+1}{2}x^{\frac{\gamma}{\gamma+1}} \tag{16}$$

Furthermore, we have $U_2(n) = \frac{(\gamma+1)^2}{2\gamma+1}n^{2\gamma+1} + O(n^{2\gamma})$, so we can write

$$\begin{aligned}
G(x) &= U_2(V_1^{-1}(x) + 1) + \mathbb{E}\left[V_2(V_1^{-1}(\mathcal{Y}(l+1))|\mathcal{Y}(l) = x\right] \\
&\underset{x \gg 1}{\sim} \frac{(\gamma+1)^2}{2\gamma+1}x^{\frac{2\gamma+1}{\gamma+1}} + \frac{(\gamma+1)^2}{2\gamma+1}\mathbb{E}\left[\mathcal{Y}(l+1)^{\frac{2\gamma+1}{\gamma+1}}|\mathcal{Y}(l) = x\right] \\
&\sim \frac{(\gamma+1)^2}{2\gamma+1}\left(x^{\frac{2\gamma+1}{\gamma+1}} + x^{\frac{2\gamma+1}{\gamma+1}}\right) = 2\frac{(\gamma+1)^2}{2\gamma+1}x^{\frac{2\gamma+1}{\gamma+1}}
\end{aligned} \tag{17}$$

following the argument that $\mathbb{E}[\mathcal{Y}(l+1)|\mathcal{Y}(l) = x] \sim x$ at leading order. Then, we note similarly to Sec. 3.1 (with $\beta = \frac{1}{2\gamma+1}$), $l = \lfloor At \rfloor$ and $Y(t) = (A\beta)^{-\gamma-1}\mathcal{Y}(\lfloor At \rfloor)$ and take the limit of $A$ large,

$$(A\beta)^{-\gamma-1}\left(\mathcal{Y}_{\lfloor At \rfloor + 1} - \mathcal{Y}_{\lfloor At \rfloor}\right) = Y\left(t + \frac{1}{A}\right) - Y(t) \tag{18}$$

$$= (A\beta)^{-\gamma-1}\left[\sqrt{G((A\beta)^{\gamma+1}Y(t))}\epsilon_{\lfloor At \rfloor} + F((A\beta)^{\gamma+1}Y(t))\right] \tag{19}$$

which leads to

$$\begin{aligned}
Y\left(t + \frac{1}{A}\right) - Y(t) &\sim (A\beta)^{-\gamma-1}\left(\sqrt{2\frac{(\gamma+1)^2}{2\gamma+1}(A\beta)^{2\gamma+1}Y^{\frac{2\gamma+1}{2\gamma+2}}(t)}\sqrt{A}dW_t + \frac{\gamma+1}{2}(A\beta)^\gamma Y^{\frac{\gamma}{\gamma+1}}(t)\right) \\
&= \sqrt{2}(\gamma+1)Y^{\frac{2\gamma+1}{2\gamma+2}}(t)dW_t + A^{-1}\frac{(\gamma+1)(2\gamma+1)}{2}Y^{\frac{\gamma}{\gamma+1}}(t)
\end{aligned} \tag{20}$$

Choosing $1/A = dt$, we then find the SDE

$$dY(t) = \frac{(\gamma+1)(2\gamma+1)}{2}Y(t)^{\frac{\gamma}{\gamma+1}}dt + \sqrt{2}(\gamma+1)Y(t)^{\frac{2\gamma+1}{2\gamma+2}}dW_t \tag{21}$$

Now, we show that we can write $Y(t) = B_1(t)^{\gamma+1}$ where $B_1(t)$ is (one half of) a BESQ$_1$, which means, in our notations

$$dB_1(t) = \frac{dt}{2} + \sqrt{2B_1(t)}dW_t \tag{22}$$

where $dW_t$ is uncorrelated, mean zero, variance 1 gaussian white noise. Using the (generalized) Itô formula, one obtains

$$\begin{aligned}
dY(t) = d(B_1^{\gamma+1}(t)) &= (\gamma+1)B_1(t)^\gamma dB_1(t) + \frac{\gamma(\gamma+1)}{2}B(t)^{\gamma-1} \times (2B_1(t)dt) \\
&= (\gamma+1)B_1(t)^\gamma\left(\frac{dt}{2} + \sqrt{2B_1(t)}dW(t)\right) + \gamma(\gamma+1)B_1(t)^\gamma dt
\end{aligned} \tag{23}$$

5

which is exactly (21), after replacing $B_1 = Y^{\frac{1}{\gamma+1}}$. Since $\mathcal{Z}(l) = V_1^{-1}(\mathcal{Y}(l))$ and $V_1^{-1}(x) \sim x^{\frac{1}{1+\gamma}}$, the limit process $Z(t) = \lim_{A \to \infty} A^{-1}\mathcal{Z}(\lfloor At \rfloor)$ is a BESQ$_1$ as well.

## 3.3 SESRW

This case is much more technical than the others. We will only give the derivation for the $\kappa = 1$ case, which is referred to as TSAW in the literature. We refer the reader to [2] for more details.

### 3.3.1 TSAW

For the TSAW, $w(n) = e^{\beta n}$ with $\beta > 0$ for attractive interaction and $\beta < 0$ for repulsive interaction.
The functions $U, V$ become

$$U_p(n) = \frac{e^{2p\beta n} - 1}{e^{2p\beta} - 1}, \quad V_p(n) = e^{p\beta} \frac{e^{2p\beta n} - 1}{e^{2p\beta} - 1} \tag{24}$$

and $V_1^{-1}(x) = \frac{\log(e^{-\beta}(e^\beta + e^{2\beta}x - x))}{2\beta} \sim \frac{\log x}{2\beta}$. One has

$$F(x) = \mathbb{E}\left[\mathcal{Y}(l+1)|\mathcal{Y}(l) = x\right] - x = U_1(V_1^{-1}(x) + 1) - x = 1 + (e^\beta - 1)x \tag{25}$$

$$G(x) = U_2(V_1^{-1}(x) + 1) + \mathbb{E}\left[V_2(V_1^{-1}(\mathcal{Y}(l+1)))|\mathcal{Y}(l) = x\right] \tag{26}$$

$$\sim \frac{e^{2\beta} - 1}{e^{2\beta} + 1}x^2 \left[e^{2\beta} + \mathbb{E}\left[\left(1 + \frac{\mathcal{Y}(l+1) - \mathcal{Y}(l)}{x}\right)^2 |\mathcal{Y}(l) = x\right]\right] \tag{27}$$

The difference $\mathcal{Y}(l+1) - \mathcal{Y}(l)$ is, by definition, of order $F(\mathcal{Y}(l))$ : given the above, the latter is itself of order $\mathcal{Y}(l)$. In the other cases of SATW and PSRW, this was not the case, as $\mathcal{Y}(l+1) - \mathcal{Y}(l)$ was much smaller than $\mathcal{Y}(l)$. However, for $x \gg 1$, $\frac{F(x)}{x} = e^\beta - 1 \sim \beta$ when $\beta \ll 1$ and in this condition we can make the expansion

$$G(x) \underset{x \gg 1, \beta \ll 1}{\sim} \frac{2\beta}{2}(e^{2\beta} + e^{2\beta})x^2 = \left(\sqrt{2\beta}e^\beta x\right)^2 \tag{28}$$

Let us write $G(x) \sim (gx)^2$. Noticing that both $F(x), \sqrt{G(x)} \propto x$ for large $x$, we have a hint that $\mathcal{Y}(l)$ is of exponential order in $l$ for large $l$, in contrast with the former PSRW, SATW cases. Thus, we define the scaling limit $y(t)$ of $y_A(t) = \frac{\log \mathcal{Y}(\lfloor At \rfloor)}{2\beta\sqrt{A}}$ as $A \to \infty$. We find (setting $A = \frac{1}{dt}$, $\epsilon_{\lfloor At \rfloor} = \frac{dW_t}{\sqrt{dt}}$ and $dW_t^2 = dt$)

$$dy_A(t) = y_A(t + dt) - y_A(t) = \frac{1}{2\beta\sqrt{A}} \log\left(\frac{\mathcal{Y}(\lfloor At \rfloor + 1)}{\mathcal{Y}(\lfloor At \rfloor)}\right)$$

$$= \frac{1}{2\beta\sqrt{A}} \log\left(1 + (e^\beta - 1) + g\epsilon_{At}\right)$$

$$= \frac{1}{2\beta\sqrt{A}} \left(\beta + \log\left(1 + \frac{g}{e^\beta}\epsilon_{At}\right)\right)$$

Now if we suppose $\beta \ll e^\beta$, then we do an expansion of log and

$$dy_A(t) = \frac{1}{2\beta\sqrt{A}} \left(\beta - \frac{1}{2}\left(\frac{g}{e^\beta}\right)^2 Adt + \frac{g}{e^\beta}dW_t\sqrt{A}\right).$$

The only way this process does not explode as $A \to \infty$ is to have $g = \sqrt{2\beta}e^{\beta}$. Then, we have that the limit process of local times is a Brownian motion with variance $\frac{1}{2\beta}$

$$dy(t) = \frac{1}{\sqrt{2\beta}}dW_t. \tag{29}$$

# 4 Plot of the splitting probability of SIRW

The variety of the shapes of splitting probabilities of SIRWs is illustrated below.

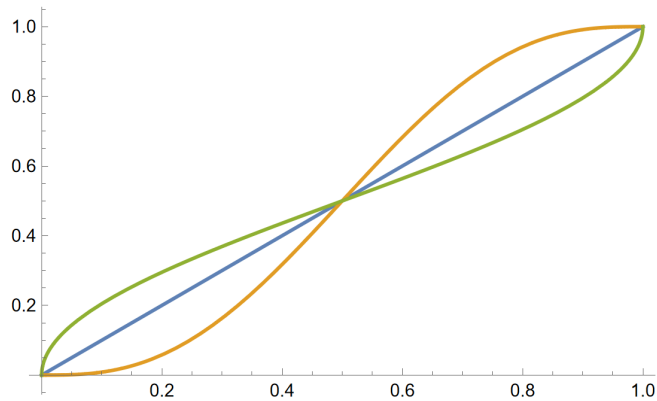

Figure 3: Plot of the splitting probability for various SIRW. Blue : Brownian motion, orange : SATW$_{\phi=3}$, green : PSRW, SESRW$_{\kappa,\beta}$, SATW$_{\phi=1/2}$.



\end{document}